\documentclass[aps,prl,twocolumn,superscriptaddress,showpacs]{revtex4}
\usepackage{color}
\usepackage{eso-pic}
\usepackage{dcolumn}% Align table columns on decimal point
\usepackage{bm}% bold math

\newif\ifpdf
 \ifx\pdfoutput\undefined
 \pdffalse                  % we are not running PDFLaTeX
\else
 \pdfoutput=1               % we are running PDFLaTeX
 \pdftrue
\fi

\ifpdf
 \usepackage{url,boxedminipage}
 \usepackage{graphicx,thumbpdf}
 \DeclareGraphicsExtensions{.pdf,.jpg,.jpeg}
\else
 \usepackage{graphicx}
 \DeclareGraphicsExtensions{.eps,.ps,.eps.gz,.ps.gz}
 \newcommand{\href}[2]{#2}                   % suppress URL's in PS files
 \usepackage[dvips,pagebackref]{hyperref}    % also create Page-back-refrences
\fi % end ifpdf

\begin{document}

%%%%%%%%%%%%%%%%%%%%%%%%%%%%%%%%%%%%%%%%%%%%%
% Uncomment this line to print the Fermilab
% preprint number
%
{\flushright FERMILAB-PUB-09-036-E}
\preprint{FERMILAB-PUB-09-036-E}
%%%%%%%%%%%%%%%%%%%%%%%%%%%%%%%%%%%%%%%%%%%%%

%%%%%%%%%%%%%%%%%%%%%%%%%%%%%%%%%%%%%%%%%%%%%
% Uncomment this line to enable line numbers
%
% \linenumbers
%%%%%%%%%%%%%%%%%%%%%%%%%%%%%%%%%%%%%%%%%%%%%

%\preprint{}

\newcommand{\Pt}{\hbox{$p_T$}}
\newcommand{\pt}{\hbox{$p_T$}}
\newcommand{\tz}{\hbox{$t_0$}}
\newcommand{\mtof}{\hbox{$m_{\rm TOF}$}}
\newcommand{\fbinv}{\hbox{fb$^{-1}$}}
\newcommand{\gevc}{\hbox{GeV/$c$}}
\newcommand{\gevcc}{\hbox{GeV/$c^2$}}
\newcommand{\etal}{\hbox{\textit{et al.}}}
\newcommand{\pbarp}{\ensuremath{\bar{p}p}}
\newcommand{\ppbar}{\ensuremath{p\bar{p}}}
\newcommand{\topsqrk}{\hbox{$\tilde{t}$}}

\title{Search for Long-Lived Massive Charged Particles in
       1.96~TeV \ppbar\ Collisions}


% \author{}
%\thanks{}
% \affiliation{}
%
% The full author list
%
\affiliation{Institute of Physics, Academia Sinica, Taipei, Taiwan 11529, Republic of China} 
\affiliation{Argonne National Laboratory, Argonne, Illinois 60439} 
\affiliation{University of Athens, 157 71 Athens, Greece} 
\affiliation{Institut de Fisica d'Altes Energies, Universitat Autonoma de Barcelona, E-08193, Bellaterra (Barcelona), Spain} 
\affiliation{Baylor University, Waco, Texas  76798} 
\affiliation{Istituto Nazionale di Fisica Nucleare Bologna, $^x$University of Bologna, I-40127 Bologna, Italy} 
\affiliation{Brandeis University, Waltham, Massachusetts 02254} 
\affiliation{University of California, Davis, Davis, California  95616} 
\affiliation{University of California, Los Angeles, Los Angeles, California  90024} 
\affiliation{University of California, San Diego, La Jolla, California  92093} 
\affiliation{University of California, Santa Barbara, Santa Barbara, California 93106} 
\affiliation{Instituto de Fisica de Cantabria, CSIC-University of Cantabria, 39005 Santander, Spain} 
\affiliation{Carnegie Mellon University, Pittsburgh, PA  15213} 
\affiliation{Enrico Fermi Institute, University of Chicago, Chicago, Illinois 60637}
\affiliation{Comenius University, 842 48 Bratislava, Slovakia; Institute of Experimental Physics, 040 01 Kosice, Slovakia} 
\affiliation{Joint Institute for Nuclear Research, RU-141980 Dubna, Russia} 
\affiliation{Duke University, Durham, North Carolina  27708} 
\affiliation{Fermi National Accelerator Laboratory, Batavia, Illinois 60510} 
\affiliation{University of Florida, Gainesville, Florida  32611} 
\affiliation{Laboratori Nazionali di Frascati, Istituto Nazionale di Fisica Nucleare, I-00044 Frascati, Italy} 
\affiliation{University of Geneva, CH-1211 Geneva 4, Switzerland} 
\affiliation{Glasgow University, Glasgow G12 8QQ, United Kingdom} 
\affiliation{Harvard University, Cambridge, Massachusetts 02138} 
\affiliation{Division of High Energy Physics, Department of Physics, University of Helsinki and Helsinki Institute of Physics, FIN-00014, Helsinki, Finland} 
\affiliation{University of Illinois, Urbana, Illinois 61801} 
\affiliation{The Johns Hopkins University, Baltimore, Maryland 21218} 
\affiliation{Institut f\"{u}r Experimentelle Kernphysik, Universit\"{a}t Karlsruhe, 76128 Karlsruhe, Germany} 
\affiliation{Center for High Energy Physics: Kyungpook National University, Daegu 702-701, Korea; Seoul National University, Seoul 151-742, Korea; Sungkyunkwan University, Suwon 440-746, Korea; Korea Institute of Science and Technology Information, Daejeon, 305-806, Korea; Chonnam National University, Gwangju, 500-757, Korea} 
\affiliation{Ernest Orlando Lawrence Berkeley National Laboratory, Berkeley, California 94720} 
\affiliation{University of Liverpool, Liverpool L69 7ZE, United Kingdom} 
\affiliation{University College London, London WC1E 6BT, United Kingdom} 
\affiliation{Centro de Investigaciones Energeticas Medioambientales y Tecnologicas, E-28040 Madrid, Spain} 
\affiliation{Massachusetts Institute of Technology, Cambridge, Massachusetts  02139} 
\affiliation{Institute of Particle Physics: McGill University, Montr\'{e}al, Qu\'{e}bec, Canada H3A~2T8; Simon Fraser University, Burnaby, British Columbia, Canada V5A~1S6; University of Toronto, Toronto, Ontario, Canada M5S~1A7; and TRIUMF, Vancouver, British Columbia, Canada V6T~2A3} 
\affiliation{University of Michigan, Ann Arbor, Michigan 48109} 
\affiliation{Michigan State University, East Lansing, Michigan  48824}
\affiliation{Institution for Theoretical and Experimental Physics, ITEP, Moscow 117259, Russia} 
\affiliation{University of New Mexico, Albuquerque, New Mexico 87131} 
\affiliation{Northwestern University, Evanston, Illinois  60208} 
\affiliation{The Ohio State University, Columbus, Ohio  43210} 
\affiliation{Okayama University, Okayama 700-8530, Japan} 
\affiliation{Osaka City University, Osaka 588, Japan} 
\affiliation{University of Oxford, Oxford OX1 3RH, United Kingdom} 
\affiliation{Istituto Nazionale di Fisica Nucleare, Sezione di Padova-Trento, $^y$University of Padova, I-35131 Padova, Italy} 
\affiliation{LPNHE, Universite Pierre et Marie Curie/IN2P3-CNRS, UMR7585, Paris, F-75252 France} 
\affiliation{University of Pennsylvania, Philadelphia, Pennsylvania 19104}
\affiliation{Istituto Nazionale di Fisica Nucleare Pisa, $^z$University of Pisa, $^{aa}$University of Siena and $^{bb}$Scuola Normale Superiore, I-56127 Pisa, Italy} 
\affiliation{University of Pittsburgh, Pittsburgh, Pennsylvania 15260} 
\affiliation{Purdue University, West Lafayette, Indiana 47907} 
\affiliation{University of Rochester, Rochester, New York 14627} 
\affiliation{The Rockefeller University, New York, New York 10021} 
\affiliation{Istituto Nazionale di Fisica Nucleare, Sezione di Roma 1, $^{cc}$Sapienza Universit\`{a} di Roma, I-00185 Roma, Italy} 

\affiliation{Rutgers University, Piscataway, New Jersey 08855} 
\affiliation{Texas A\&M University, College Station, Texas 77843} 
\affiliation{Istituto Nazionale di Fisica Nucleare Trieste/Udine, I-34100 Trieste, $^{dd}$University of Trieste/Udine, I-33100 Udine, Italy} 
\affiliation{University of Tsukuba, Tsukuba, Ibaraki 305, Japan} 
\affiliation{Tufts University, Medford, Massachusetts 02155} 
\affiliation{Waseda University, Tokyo 169, Japan} 
\affiliation{Wayne State University, Detroit, Michigan  48201} 
\affiliation{University of Wisconsin, Madison, Wisconsin 53706} 
\affiliation{Yale University, New Haven, Connecticut 06520} 
\author{T.~Aaltonen}
\affiliation{Division of High Energy Physics, Department of Physics, University of Helsinki and Helsinki Institute of Physics, FIN-00014, Helsinki, Finland}
\author{J.~Adelman}
\affiliation{Enrico Fermi Institute, University of Chicago, Chicago, Illinois 60637}
\author{T.~Akimoto}
\affiliation{University of Tsukuba, Tsukuba, Ibaraki 305, Japan}
\author{B.~\'{A}lvarez~Gonz\'{a}lez$^s$}
\affiliation{Instituto de Fisica de Cantabria, CSIC-University of Cantabria, 39005 Santander, Spain}
\author{S.~Amerio$^y$}
\affiliation{Istituto Nazionale di Fisica Nucleare, Sezione di Padova-Trento, $^y$University of Padova, I-35131 Padova, Italy} 

\author{D.~Amidei}
\affiliation{University of Michigan, Ann Arbor, Michigan 48109}
\author{A.~Anastassov}
\affiliation{Northwestern University, Evanston, Illinois  60208}
\author{A.~Annovi}
\affiliation{Laboratori Nazionali di Frascati, Istituto Nazionale di Fisica Nucleare, I-00044 Frascati, Italy}
\author{J.~Antos}
\affiliation{Comenius University, 842 48 Bratislava, Slovakia; Institute of Experimental Physics, 040 01 Kosice, Slovakia}
\author{G.~Apollinari}
\affiliation{Fermi National Accelerator Laboratory, Batavia, Illinois 60510}
% ...

\collaboration{CDF Collaboration\footnote{With visitors from $^a$University of Massachusetts Amherst, Amherst, Massachusetts 01003,
$^b$Universiteit Antwerpen, B-2610 Antwerp, Belgium, 
$^c$University of Bristol, Bristol BS8 1TL, United Kingdom,
$^d$Chinese Academy of Sciences, Beijing 100864, China, 
$^e$Istituto Nazionale di Fisica Nucleare, Sezione di Cagliari, 09042 Monserrato (Cagliari), Italy,
$^f$University of California Irvine, Irvine, CA  92697, 
$^g$University of California Santa Cruz, Santa Cruz, CA  95064, 
$^h$Cornell University, Ithaca, NY  14853, 
$^i$University of Cyprus, Nicosia CY-1678, Cyprus, 
$^j$University College Dublin, Dublin 4, Ireland,
$^k$University of Edinburgh, Edinburgh EH9 3JZ, United Kingdom, 
$^l$University of Fukui, Fukui City, Fukui Prefecture, Japan 910-0017
$^m$Kinki University, Higashi-Osaka City, Japan 577-8502
$^n$Universidad Iberoamericana, Mexico D.F., Mexico,
$^o$Queen Mary, University of London, London, E1 4NS, England,
$^p$University of Manchester, Manchester M13 9PL, England, 
$^q$Nagasaki Institute of Applied Science, Nagasaki, Japan, 
$^r$University of Notre Dame, Notre Dame, IN 46556,
$^s$University de Oviedo, E-33007 Oviedo, Spain, 
$^t$Texas Tech University, Lubbock, TX  79609, 
$^u$IFIC(CSIC-Universitat de Valencia), 46071 Valencia, Spain,
$^v$University of Virginia, Charlottesville, VA  22904,
$^w$Bergische Universit\"at Wuppertal, 42097 Wuppertal, Germany,
$^{ee}$On leave from J.~Stefan Institute, Ljubljana, Slovenia, 
}}
\noaffiliation

%
% The following 2 lines are not needed when the full author list is included:
%
% \collaboration{CDF Collaboration}
% \noaffiliation


%\date{\today}

\begin{abstract}
  We performed a signature-based search for long-lived  charged massive
  particles (CHAMPs) produced in 1.0~$\rm{fb}^{-1}$ of $p\overline p$ collisions
  at $\sqrt{s}=1.96\;{\rm TeV}$, collected with the CDF~II 
  detector using a high transverse-momentum (\pt) muon trigger. The search used 
  time-of-flight 
  to isolate slowly moving, high-\pt\ particles. One event passed our selection
  cuts with an expected background of $1.9 \pm 0.2$ events.
  We set an upper bound on the production cross section, and,
  interpreting this result within the context of
  a stable scalar top-quark model, set a lower
  limit on the particle mass of 249~\gevcc\ at 95\% C.L.
  %%%%  was 587 characters , now 599(?)
  %%%%%%%%%%%%%%%%%%%%%%%%%% 
\end{abstract}

%\keywords{}

\maketitle


Most searches for massive particles arising from physics beyond the standard 
model (SM)
rely upon the assumption that the
particles decay immediately. 
Long-lived or stable non-SM states could exist, however, due to 
a new symmetry \cite{new-sym-models}, a weak coupling 
\cite{weak-coupling-models}, a kinematic constraint
\cite{kin-constraint-models}, or a potential barrier
\cite{pot-barrier-models}. If the
lifetime is long compared to the
transit time through the detector, then the particle may escape the
detector, thereby evading the limits imposed by direct searches for decay
products.  However, a charged, massive long-lived particle (CHAMP) will be 
directly observable within the detector through the distinctive  signature
of a slowly moving,  high transverse-momentum  (\Pt) particle. The low
velocity results in a long time-of-flight (TOF) and an anomalously large 
ionization-energy loss rate ($dE/dx$). Since the
particle loses energy primarily through low-momentum-transfer interactions,
even if strongly interacting 
\cite{drees,smp-review}, it will be highly penetrating and will
likely be reconstructed as a muon.

Previous CHAMP search results
have been presented within the context of a variety of 
models~\cite{cdfsearch,lepsearch,lepslepton,oceanSearches}.  
CDF in Run~I, for instance, used $dE/dx$ and 
set 95\% C.L.\ lower mass limits on stable fourth-generation down-type
(190~\gevcc) and up-type (220~\gevcc)  quarks~\cite{cdfsearch}. The ALEPH
experiment also used $dE/dx$ to exclude a stable scalar top squark
(\topsqrk), the supersymmetric partner of the top quark,   with a mass
below 95~\gevcc~at 95\% C.L.~\cite{lepsearch} A
combined result from the LEP2 experiments excluded a stable supersymmetric
partner for SM leptons with a mass below 99.5~\gevcc\ at 
95\% C.L.~\cite{lepslepton} 

In this Letter, we present a blind signature-based search for isolated
CHAMPs promptly produced in $p\bar{p}$ collisions at  $\sqrt{s} = 1.96$~TeV
with the CDF~II detector~\cite{CDF} at the Fermilab Tevatron.  
Using an integrated luminosity of 1.0~\fbinv\ of \pbarp\ collisions
collected with
a high~\Pt\ muon trigger, the analysis isolated CHAMP candidates by 
calculating their mass from their measured velocity and momentum.
We interpret the results within two scenarios. The first case, production of
a single CHAMP within a reference volume
of the CDF~II detector, is largely
model independent. The second scenario assumes a benchmark
model for
stable top-squark-pair production within the reference volume. 
Since the leading-order contributions 
to \topsqrk\ 
production depend only upon the \topsqrk\  mass \cite{stop-prod}, the 
result will
generally apply to all stable \topsqrk\ production models.

Details of the CDF~II detector can be found in Ref.~\cite{CDF}.
CDF measures the trajectories and momenta of charged 
particles using an inner silicon-strip detector \cite{silicon} and an
open-cell drift chamber (COT)~\cite{cot}.
A TOF detector~\cite{tof}
surrounding the outer tracker allows precise arrival time measurements for
tracks projected into the detector with a pseudorapidity \cite{eta_defn}
in the range $|\eta| \lesssim 1$. Calorimeters located outside the
tracking volume measure energy deposition of  particles, and 
prevent all but the most penetrating  from reaching the muon
detectors~\cite{muon-det} positioned beyond the calorimeters. 

Our data sample was collected with a trigger that
identifies muon candidates with $|\eta| < 0.7$ and $\Pt > 18$~\gevc. 
An event entered the analysis if the highest-\Pt\ muon candidate
reconstructed offline had  $\Pt > 20$~\gevc, 
originated from the
most energetic \pbarp\ collision,
passed quality
criteria that reduce backgrounds from punch-through and particles that decay 
in-flight, and satisfied a calorimeter energy isolation criterion in which 
the ratio $\Sigma E_T(0.4) / \Pt({\rm muon}) < 0.1$, where $\Sigma E_T(0.4)$ 
is the sum of transverse energy within a cone of $\Delta R = 0.4$ around
the  candidate's direction, excluding the energy deposited by the candidate
itself.

We assign the selected events to signal or control sub-samples depending upon
whether the track of the highest-\Pt\ muon candidate is a signal-region 
($\Pt > 40$~\gevc)
or control-region ($20 < \Pt < 40$~\gevc) track.
The second-highest-\Pt\ muon candidate
(or the highest-\Pt\ non-muon track in events with only one muon candidate)
is also a signal- or control-region track if it is in the same \Pt\ region 
and originates from the same vertex as the first muon candidate.
Tracks  with $\Pt < 20$~\gevc\ 
are used to measure the \ppbar\ interaction time (\tz)
and are referred to as ``\tz\ tracks".  The
event \tz, which is needed to determine the velocity of signal- and 
control-region tracks, is estimated using a maximum likelihood fit to all 
\tz\ tracks from an interaction vertex,
simultaneously taking into account all possible mass hypotheses.   The \tz\
resolution of single tracks is about 120 ps, so a single \tz\ track is 
adequate to obtain the interaction time.

To separate a CHAMP signal from background, we use the velocity and
momentum to calculate the mass of the candidate particle. In events with
two signal-region or control-region tracks, both are considered. The
track velocity for all candidate and control-region tracks is measured by 
dividing
the path  length of the track by its TOF.  
The measured average velocity, $\beta=v/c$, and single-track resolution 
of control-region tracks is 
$1.000 \pm 0.029$, but with significant non-Gaussian tails.
For signal-region tracks, we require $\beta < 0.9$ to
suppress SM particles.

The non-Gaussian tails in the time resolution functions introduce a large
background to the CHAMP candidate sample.
The residuals to the track fit in the COT can be used 
to estimate the \tz\
and track $\beta$  with
resolutions that are about a factor of three worse than those made with the 
TOF  detector, but that are
reliably parameterized by single Gaussian distributions. Requiring that
the event \tz\ and candidate track $\beta$ measurements from the 
TOF detector and COT agree reduces this background.

Cosmic-ray muons are uncorrelated in time with $p\bar{p}$ 
interactions and
present a potentially serious background.  
In a sample of $1.5\times 10^5$ cosmic rays, only four pass the
CHAMP selection. After applying a cosmic-ray filter \cite{cosmicID}, we
expect negligible residual cosmic-ray background. The filter
removes less than 1\% of signal events.

We estimate the efficiency for identifying a CHAMP candidate within our  two
scenarios.  In
general, CHAMPs are expected to have very large \Pt\ and be highly 
isolated.
Final-state radiation is strongly suppressed, 
even if the CHAMP is strongly interacting \cite{drees}. 
These 
characteristics make $W\rightarrow l\nu$ and 
$Z\rightarrow l^+l^-$ events, where $l$ is either an electron or muon,  
reasonable models for both the isolated CHAMP
track and the underlying event. 

We use the muons in $Z\rightarrow \mu^+\mu^-$ events selected from the original
trigger sample to measure the trigger
and track reconstruction efficiency for a single 
muon to be ($94.0 \pm 0.3$)\%.  
To study the $\beta$ dependence of the tracking efficiency,
we isolate slow deuterons and pions using $dE/dx$ in the
tracking detector and measure the ratio of deuterons to pions, which we assume
is constant as a function of $\beta$.
We find that the efficiency 
is constant for $\beta > 0.4$ and drops for slower particles, a result 
confirmed in a CHAMP Monte Carlo
simulation (MC) \cite{pythia}. We therefore assume a flat efficiency of 
$(94.0 \pm 0.3)$\% for $\beta > 0.4$ and zero for $\beta < 0.4$ for CHAMPs.

Using vertices and electron tracks in
$W\rightarrow e\nu$ events, we determine the efficiency for
finding the primary event vertex, calculating an event \tz, and
reconstructing an isolated CHAMP track from the
vertex to be ($71.4 \pm 0.2$)\%. The event \tz\  and 
track-vertex association dominate the losses in this efficiency (87\% and
86\% respectively).

The efficiency for measuring the arrival time in the TOF detector  for CHAMP
tracks that are within 
the muon detector acceptance is determined directly from
the muon data; for tracks that are not within 
the muon detector's acceptance, we
use electron tracks in $W\rightarrow e\nu$ events. 
Including the efficiency for the TOF result to be consistent
with COT timing information, we obtain a TOF measurement efficiency 
of ($62.8 \pm
2.6$)\%  for tracks within 
the muon detectors and ($56.3 \pm 2.7$)\% 
for other tracks. The criteria used to identify well-measured arrival times
account for most of the efficiency loss.

The dominant systematic uncertainties in the efficiencies %%%
are a 5\% value to cover the effect of errors in the modeling of initial  
and final state radiation and track multiplicities in CHAMP events on the 
vertex and \tz\ efficiencies, 
and a 3\% uncertainty 
in the arrival time efficiency to cover differences observed for
electrons, muons, and changes in the TOF detector gain during the run.

Strongly interacting CHAMPs are subject to QCD effects 
\cite{drees,smp-review} that can
reduce the overall detection efficiency relative to that of weakly interacting
CHAMPs. 
Quark-like CHAMPs, for instance, can
hadronize into either charged or neutral color-singlet states. Charge-exchange
interactions in the material of the detector can change an initially charged
particle into a neutral particle, and visa versa, before it reaches the muon 
detectors.
At least one CHAMP must leave a track segment in both the COT and the muon 
chambers to satisfy our trigger.

In order to estimate the efficiency loss due to these hadronic effects, we
consider  the case of an up-quark-like CHAMP, $Q$, that hadronizes into a
$Q\bar{q}$ or  $\bar{Q}q$ R-hadron state \cite{rhadron}. The fraction
hadronizing into a charged R-hadron is assumed to be (52.9$\pm$2.9)\%, 
based upon the rate for charged $b$-meson production measured at LEP
\cite{lepflavor}.
The center-of-mass energy for
collisions between a massive $Q$ moving at low velocity and a light quark
is small. As a result, hadronic interactions of the R-hadron with
the detector material involve primarily the light quark while the
$Q$ remains a spectator \cite{drees,smp-review}. 
Since the R-hadron contains a single light valence quark, we assume the 
interaction length for the
R-hadron to be three times that for a proton.
Under these assumptions, 
we estimate that the probability that an initially charged
R-hadron undergoes re-hadronization before reaching the outer-most
of the two layers of muon detectors is 93\%.
At each interaction, the $Q$
re-hadronizes according to the same prescription as for the initial
hadronization. To estimate the systematic uncertainty, we take the
difference between the result above and the efficiency assuming that 100\%
of R-hadrons re-hadronize.

Combining all efficiencies, the net efficiency for 
detecting a single, weakly interacting CHAMP within the muon trigger 
acceptance is $(38\pm 2)$\%;  for
a strongly interacting up-quark-like CHAMP, the efficiency is $(8.8\pm 1.6)$\%.

As a reference model we use {\sc Pythia} \cite{pythia} to calculate the
geometric  and kinematic acceptance for top-squark pair production. 
The trigger and detection efficiencies are calculated by combining
the single-track and vertex-finding efficiencies as estimated for the
case of a single up-quark-like CHAMP with the relative rate
at which one or two top-squark R-hadrons are within the fiducial volume
of the detector as predicted by the MC.  
The acceptances for various \topsqrk\ masses are listed 
in Table~\ref{tab:results}.

\begin{figure}
\begin{center}
  \includegraphics[width=\columnwidth]{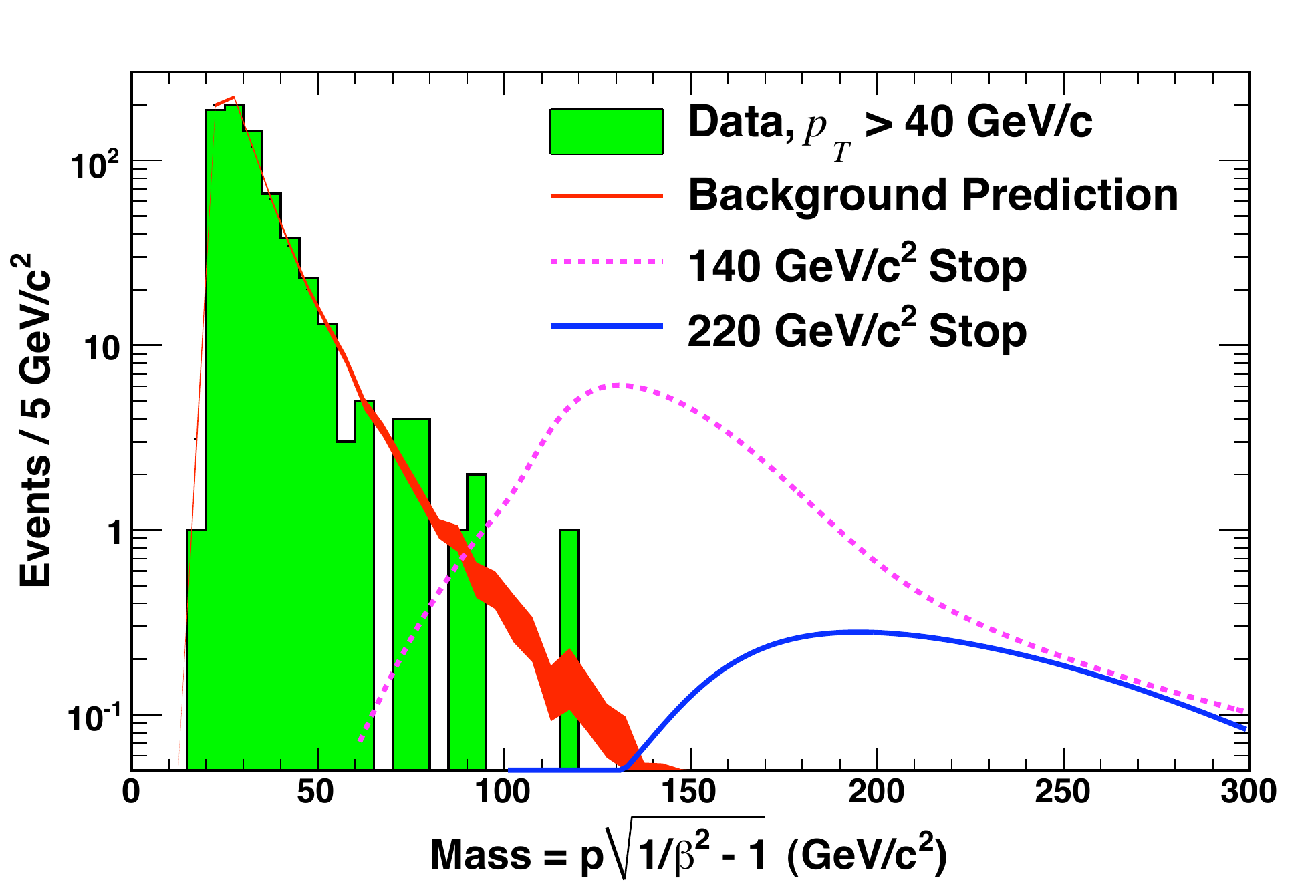}
  \caption{ Observed (histogram) and predicted (band) %TOF 
            mass distributions for
	    candidate tracks in the muon sample. The curves on the right
            show the MC distributions expected for a 140 and a 220~\gevcc\ 
	    long-lived stop.}
  \label{fig:muon_sig_tof_mass}
\end{center}
\end{figure}

Figure \ref{fig:muon_sig_tof_mass} shows the observed and predicted
mass distribution for tracks in the signal region. 
The uncertainty in the $\beta$ measurement is independent of the
momentum for tracks with $\beta \approx 1$. We therefore obtain an absolute 
prediction for the background mass
distribution for a given set of tracks 
by convolving the momentum
distribution for those tracks with the distribution of $\sqrt{1/\beta^2 - 1}$, 
normalized to unit area, for control-region tracks.
We find agreement between the observed and predicted 
mass distributions within the control and signal-region electron tracks and 
within
the control region of the muon sample.  
The background prediction for the signal region is shown by the band in 
Fig.~\ref{fig:muon_sig_tof_mass}.

We find one candidate track with a 
mass above 100~\gevcc\ and none above 120~\gevcc, consistent with the predicted
background of $1.9 \pm 0.2$ events above 100~\gevcc. 
From this result, we set a model-independent upper limit on the
production cross section for a single, isolated, weakly interacting CHAMP
within the muon trigger acceptance (approximately $|\eta| < 0.7$) with  $\Pt >
40$~\gevc, $0.4 < \beta < 0.9$, and a measured mass $m> 100$~\gevcc\ 
to be $\sigma < 10$~fb at 95\%~C.L.
Similarly, the cross-section limit for a  up-quark-like CHAMP under the
same assumptions is $\sigma < 48$~fb at 95\%~C.L.

\begin{figure}
\begin{center}
  \includegraphics[width=\columnwidth]{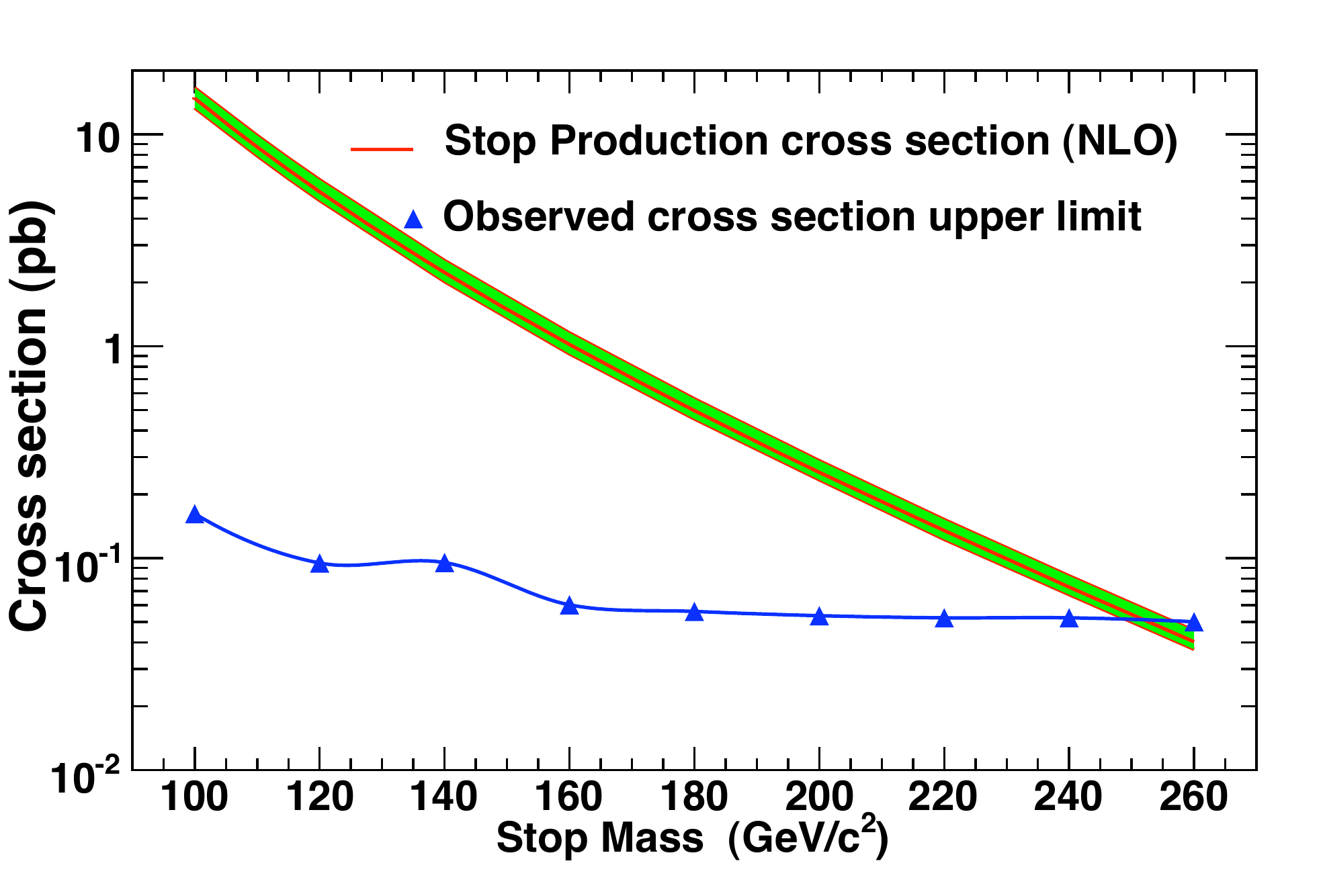}
  \caption{ The observed 95\% C.L. limits on the cross section for 
            production
	    of a stable top-squark pair (points), compared to the theoretical 
	    NLO cross section \cite{stop-prod} (curve). 
	    %as calculated by the Prospino2 program 
	    %shown as a band.
	    The band represents theoretical and parton distribution 
	    function uncertainties.
	    The intersection of the band with the limit curve yields a 
	    lower mass limit for a stable top squark of 249~\gevcc. }
  \label{fig:StopXSecLimitCMUP}
\end{center}
\end{figure}

To count the number of events consistent with a stable \topsqrk\ of a given 
mass $m_s$, we must take into account our mass resolution.
For tracks with $\beta > 0.4$ and 
momenta in the signal region, the mass resolution is determined by the
momentum resolution \cite{COT-pt-resoln}, which is well modeled
by the MC. We can therefore accurately predict the \topsqrk\ mass line shape.
We search for a \topsqrk\ signal by integrating
all events within a one-sided window from $0.8 m_s$
upward. Table~\ref{tab:results} shows the resulting number of  
events as a function of the \topsqrk\ mass. From the estimated efficiencies 
and the number of observed events, we calculate the  95\%~C.L. upper
limit on the cross
section shown in Fig.~\ref{fig:StopXSecLimitCMUP}. The band represents the 
theoretical NLO
\topsqrk\ pair production cross section, as calculated using the 
{\sc Prospino2} program \cite{prospino2}. From the intersection 
of the edge of the band and the limit curve,
we infer a 249~\gevcc\ 95\%~C.L. lower limit on the mass of a stable 
\topsqrk.  This is the most stringent limit to date.

\begingroup
\squeezetable
\begin{table*}[bpht]
\begin{ruledtabular}
\begin{center}
  \caption{ Results of the search for stable top squarks in 1.0~\fbinv\ of
            $p\bar{p}$ collisions, as a function of the \topsqrk\ mass. }
  \begin{tabular}{ l | c c c c c c c c c }
  \topsqrk\ mass (\gevcc) & 100 & 120 & 140 & 160 & 180 & 200 & 220 & 240 & 260 \\
  \hline
  Expected background & 4.7$\pm$0.3  &	1.9$\pm$0.2  &	0.8$\pm$0.1 & 	0.37$\pm$0.05  & 0.18$\pm$0.03  & 0.09$\pm$0.02  & 0.05$\pm$0.01  & 0.03$\pm$0.01 & 0.016$\pm$0.005 \\
  Observed events     &	4 &	1 &	1 &	0 &	0 &	0 &	0 &	0 &	0 \\
  Total acceptance (\%) & 3.6$\pm$0.5 & 4.2$\pm$0.5 & 4.5$\pm$0.6 & 5.1$\pm$0.7 & 5.5$\pm$0.8 & 5.8$\pm$0.8 & 5.9$\pm$0.9 & 5.9$\pm$0.8 & 6.2$\pm$0.9 \\
  Expected limit (fb) &	190 &  120 &	90 &	71 &	61 &	56 &	55 &	53 &	51 \\
  95\% C.L. limit (fb)  &	160 &	90 &	100 &	60 &	56 &	53 &	52 &	52 &	50 \\
  \end{tabular}
  \label{tab:results}
\end{center}
\end{ruledtabular}
\end{table*}
\endgroup

In conclusion, we have used the CDF~II TOF and COT  
systems to measure the masses of highly penetrating, high-\Pt\ 
tracks.  The
observed mass distribution is consistent with the expected background, which
is dominated by  SM particles with
mis-measured velocity or momentum.  From this result,
we set upper
limits for the production cross section times acceptance of single weakly
(up-quark-like strongly) 
interacting CHAMPs to be less than 10 (48)~fb at 95\% C.L.  The 
95\% C.L. lower limit on the mass of a stable top squark is 249~\gevcc.

\begin{acknowledgments}
We thank the Fermilab staff and the technical staffs of the participating 
institutions for their vital contributions.  In particular we thank Steve 
Mrenna for help with the CHAMP simulation. This work was supported by the 
U.S. Department of Energy and National Science Foundation; the Italian 
Istituto Nazionale di Fisica Nucleare; the Ministry of Education, Culture, 
Sports, Science and Technology of Japan; the Natural Sciences and Engineering 
Research Council of Canada; the National Science Council of the Republic of 
China; the Swiss National Science Foundation; the A.P. Sloan Foundation; the 
Bundesministerium f\"ur Bildung und Forschung, Germany; the Korean Science 
and Engineering Foundation and the Korean Research Foundation; the Science 
and Technology Facilities Council and the Royal Society, UK; the Institut 
National de Physique Nucleaire et Physique des Particules/CNRS; the Russian 
Foundation for Basic Research; the Ministerio de Ciencia e Innovaci\'{o}n, and 
Programa Consolider-Ingenio 2010, Spain; the Slovak R\&D Agency; and the 
Academy of Finland.
\end{acknowledgments}


\begin{thebibliography}{99}
\bibitem{eta_defn}
  We use a coordinate system where $\phi$ is the
  azimuthal angle around the beam axis and $\theta$ is the polar angle 
  measured with respect to the proton direction.
  The pseudo-rapidity, $\eta$, is defined by $\eta \equiv -\log\tan(\theta/2)$ 
  and $\Delta R \equiv \sqrt{ \Delta\eta^2 + \Delta\phi^2 }$.
\end{thebibliography}

\cleardoublepage

\end{document}